\documentclass[twocolumn,showpacs,preprintnumbers,amsmath,amssymb]{revtex4}

\usepackage{graphicx}
\usepackage{dcolumn}
\usepackage{bm}
\begin{document}

\title{The electric pulses induced multi-resistance states in the hysteresis temperature range of
1\emph{T}-TaS$_{2}$ and 1\emph{T}-TaS$_{1.6}$Se$_{0.4}$}
\author{Yongchang Ma$^{1}$}
\altaffiliation [Email: ]{ycma@tjut.edu.cn}
\author{Dong Wu$^{2}$}
\author{Cuimin Lu$^{1}$}
\author{Cedomir Petrovic$^{3}$}
\affiliation{$^{1}$ School of Materials Science and Engineering,
Tianjin University of Technology, Tianjin 300384, China\\
$^{2}$ International Center for Quantum Materials, School of
Physics, Peking University, Beijing 100871, China\\
$^{3}$ Condensed Matter Physics and Materials Science Department,
Brookhaven National Laboratory, Upton, New York 11973, USA}

\date{\today}

\begin{abstract}
The electric pulses induced responses of 1\emph{T}-TaS$_{2}$ and
1\emph{T}-TaS$_{1.6}$Se$_{0.4}$ crystals in the commensurate
charge-density-wave phase in hysteresis temperature range have
been investigated. We observed that abrupt multi steps of the
resistance excited by electric pulses at a fixed temperature
forming multi meta-stable like states. We propose that the
response of the system corresponds to the rearrangements of the
textures of CCDW domains, and the multi-resistance states or the
nonvolatile resistance properties excited simply by electric
pulses have profound significance for the explorations of
solid-state devices.

\end{abstract}

\pacs{71.27.+a, 71.45.Lr, 71.30.+h}
\maketitle


Solid-state electronic devices (SSED) attract much attention in
scientific research and industry fields, e.g. relays, memories,
batteries and switches, for their high endurance, low noise and
space saving when compared to mechanical ones. In order to improve
functionalities, great efforts were made to understand
correlated-electron systems \cite{Imada}. Under external electric
fields of intense laser pulses or high electrical currents,
correlated electrons can exhibit anomalous nonlinear response in
transition metal compounds
\cite{Yoshida,Hollander,Vaskivskyi,Vaskivskyi2}, and more
interesting, in some systems the current-induced state can be
maintained when the electric field or the pulse is turned off.
Such $^\backprime$nonvolatile$^\prime$ behavior of the
electrically induced switching is highly desired for data-storage
applications. Of high interest is control of multi resistance
states between OFF [high resistance (HR)] and ON [low resistance
(LR)] by electric pulses \cite{Kim,Huang,Prakash,Nian,Rozenberg}.

1\emph{T}-TaS$_{2}$ is one of the classical layered
transition-metal dichalcogenides, which capture high interest in
recent years: the ultra-fast resistance switching
\cite{Stojchevska,Vaskivskyi}, the supercooled nearly commensurate
charge-density-wave (CDW) phase \cite{Yoshida}, photosensitivity
from visible to terahertz at room temperature \cite{DongWu},
electrically driven reversible insulator-metal phase transition
\cite{Yoshida,Hollander,Ritschel}. At room temperature,
1\emph{T}-TaS$_{2}$ hosts the nearly commensurate (NC) CDW.
Scanning tunnelling microscopy has revealed that the NCCDW phase
consists of trigonally packed CCDW domains separated by metallic
regions that are not fully distorted \cite{Thomson}. With
decreasing temperature, a commensurate (C) CDW phase appears below
180 K, whereas in heating the CCDW state is stable up to 220 K,
exhibiting remarkable hysteretic characteristics in various
temperature dependent properties
\cite{Yongchang_2019,DiSalvo,Wilson,Dardel,Gasparov,Tsen}.

Because of the close energy proximity of the various competing
charge ordered phases, several external perturbations or
excitations can effectively modulate the CCDW phase
\cite{Stojchevska,Yoshida,Sipos,Yu,Ludwiczak}. These include
sample dimensions, doping, thermal history, photo excitation,
electric field and pressure, raising interesting questions about
the dynamics especially near the phase transitions. In thin flake
devices, the current induced effects were discovered on the order
of kilovolts per centimeter \cite{Yoshida,Hollander,Tsen}.
However, in hysteresis region of the dc transport properties for
macroscopic specimens, electric pulses with field strength of
several tens of V/cm can induce persistent meta-stable states,
suggested to be a mixture of the high- and low-resistance
components in bulk 1\emph{T}-TaS$_{2}$ \cite{Yongchang_2019}.
Clearly, there should exist many different meta-stable or mixture
states caused by the current-induced modification of the patterns
of CCDW domains. Further, as Se substitution shows a widened
hysteretic temperature range (140 - 260 K) in charge transport
properties \cite{DiSalvo,Endo}, one would expect that the multi
meta-stable states or the nonvolatile resistance properties can be
used close to room temperature. Consequently, investigations of
the nonvolatile resistance states excited simply by electric
pulses may have significance for solid-state rheostats to
substitute the rotary or linear types and preset resistors, which
are operated mechanically and manually.

In this article, we systematically investigated the electric
pulses induced responses of the CCDW phase of 1\emph{T}-TaS$_{2}$
and 1\emph{T}-TaS$_{1.6}$Se$_{0.4}$ in the hysteresis temperature
range. The pulses could drive each of layered dichalcogenides to
multi-resistance states and maintain the values until the next
electric pulse, exhibiting nonvolatile resistance properties. The
hysteresis region of 1\emph{T}-TaS$_{1.6}$Se$_{0.4}$ is remarkably
larger than the pristine 1\emph{T}-TaS$_{2}$, and thus widens the
temperature scope for potential usage of the electronic devices.
We propose a model to explain the emergence of the nonvolatile
resistance behavior activated electrically.

The growth of 1\emph{T}-TaS$_{2}$ and
1\emph{T}-TaS$_{1.6}$Se$_{0.4}$ single crystals could be found
elsewhere \cite{DiSalvo,Yongchang}. In our experiments, we focused
on the in-plane transport properties of the sample with surface
area $1.0\times0.2$ mm$^2$ and thickness (along c-axis) about 30
$\mu$m, and the distance between the two potential electrodes is
about 0.80 mm. After the sample was transfered on a sapphire
substrate, it was fixed mechanically and connected electrically by
silver paint. In the CCDW state, the contact resistance is less
than 1\% of the bulk resistance (50 $\sim$ 100 Ohm), providing
reliable measurement results. The experimental temperature were
monitored by a Cryo$\cdot$con 32 controller with stability better
than 0.01 K. To avoid Joule heating, an electric single pulse with
the duration of 40 $\mu$s was applied on the specimen, followed by
the resistance measurement though a dc current $I_{0}$= 0.2 mA
using a Keithley 2400 source meter. Similar experimental data were
reproduced well for other samples.

\begin{figure}[t]
\begin{minipage}[t]{0.48\linewidth}
\centering
\includegraphics[width=\textwidth]{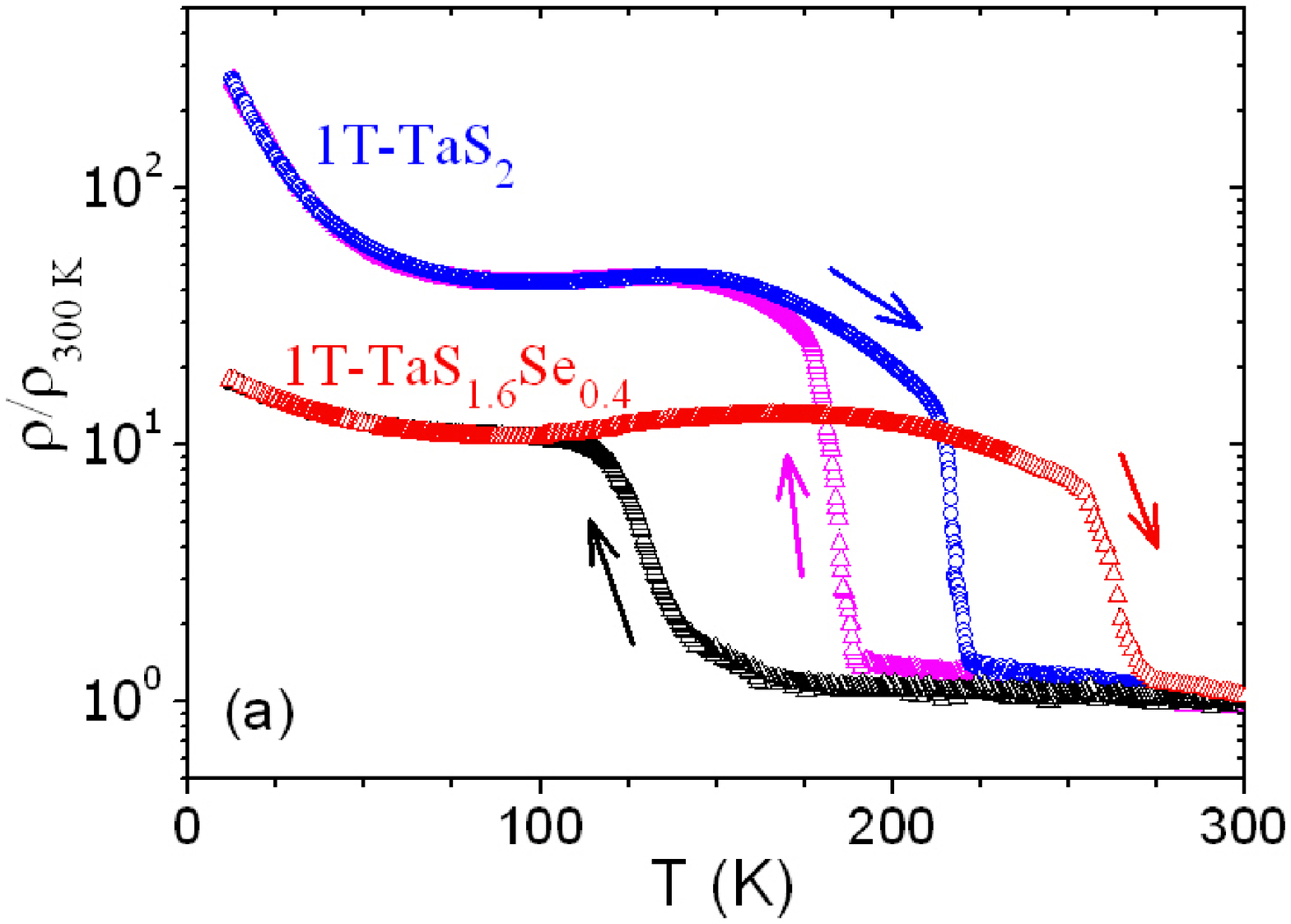}
\end{minipage}
\begin{minipage}[t]{0.48\linewidth}
\centering
\includegraphics[width=\textwidth]{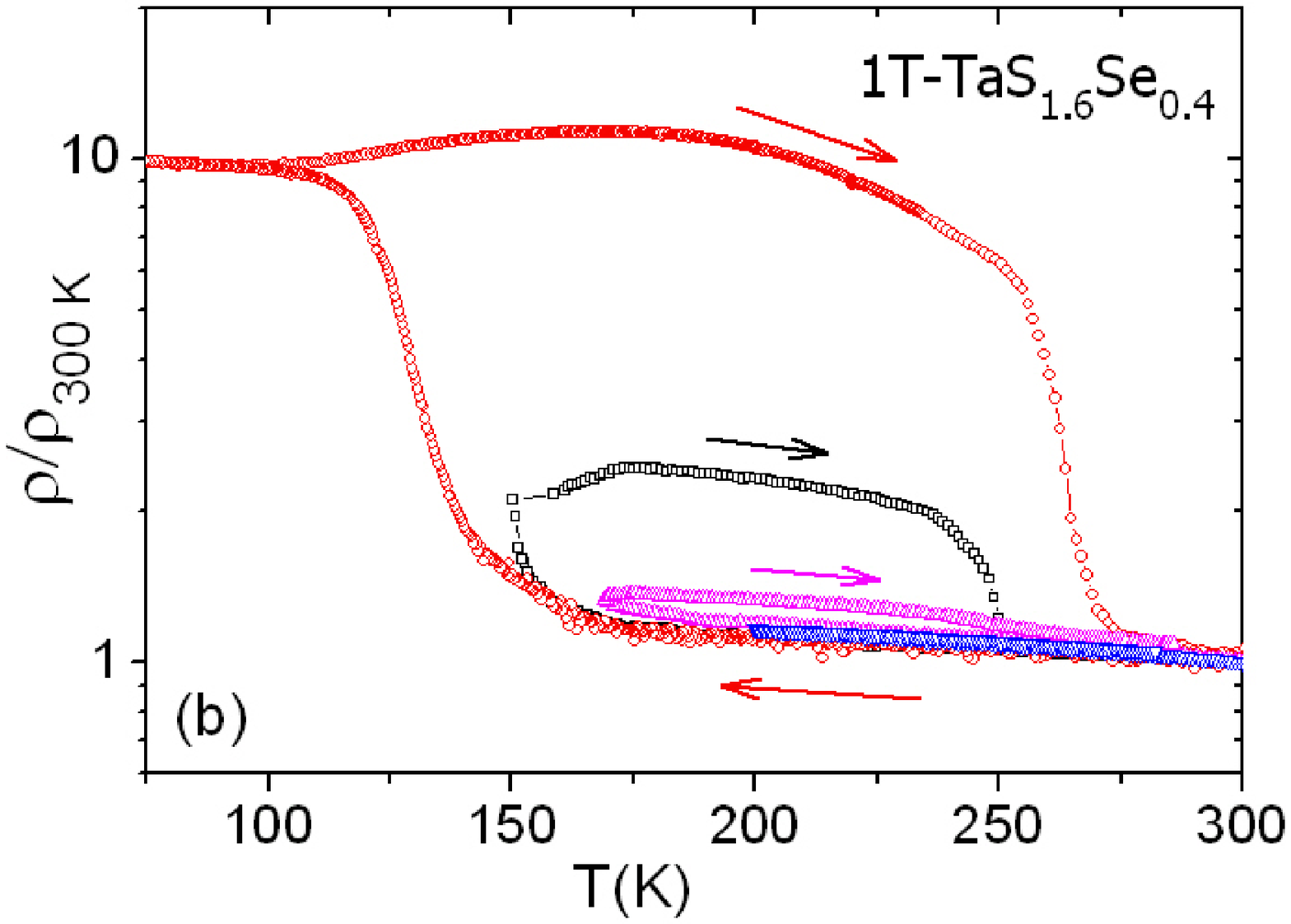}
\end{minipage}
\begin{minipage}[t]{0.48\linewidth}
\centering
\includegraphics[width=\textwidth]{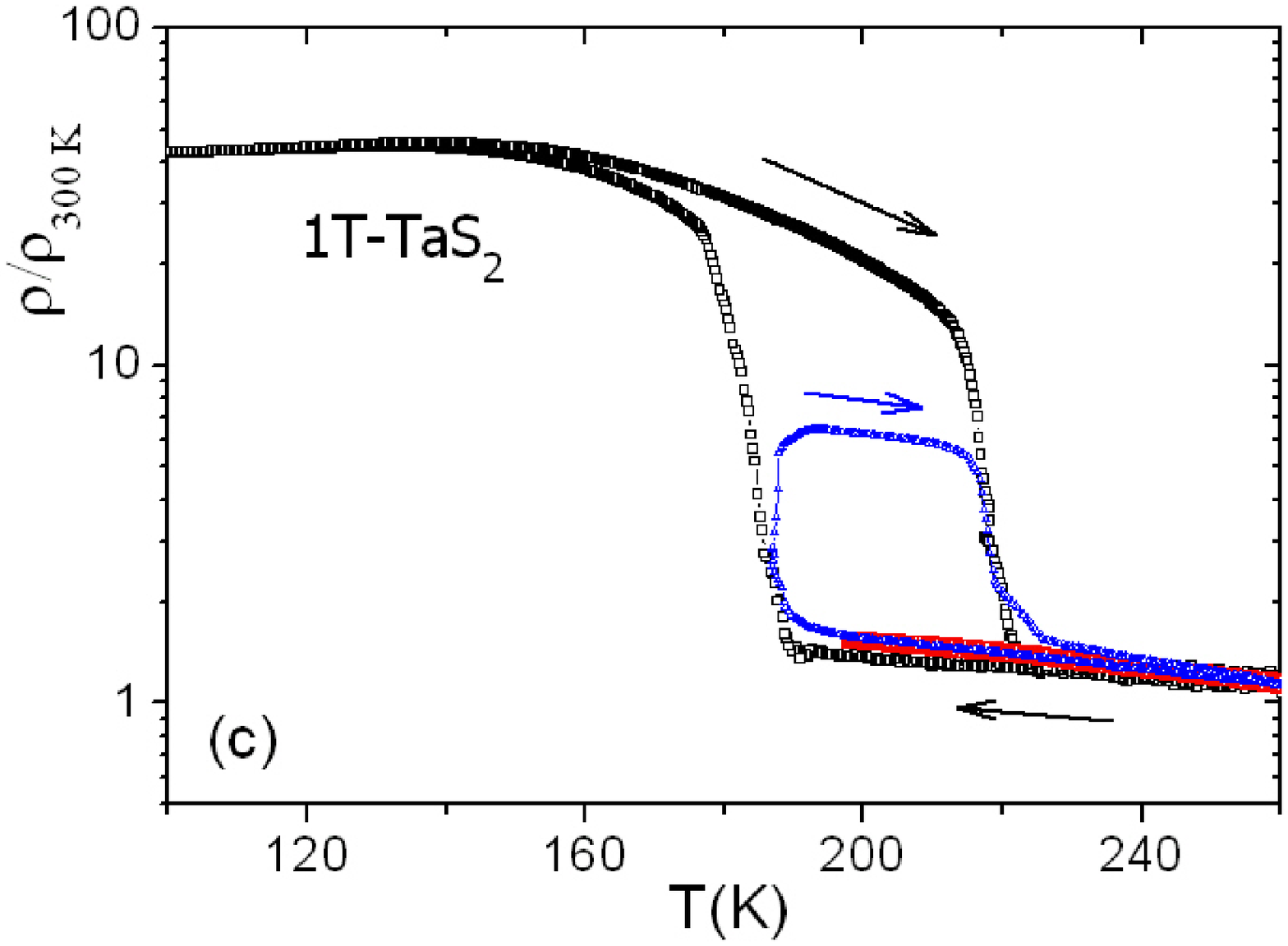}
\end{minipage}
\caption{\label{Fig01-IV-Vt} (Color online) (a) The temperature
dependent resistance (normalized to 300 K) in the \emph{ab}-plane of
1\emph{T}-TaS$_{2}$ and 1\emph{T}-TaS$_{1.6}$Se$_{0.4}$ crystals. The
arrows indicate the cooling or heating directions of the measurements.
Note the widened hysteresis region of 1\emph{T}-TaS$_{1.6}$Se$_{0.4}$
compared to 1\emph{T}-TaS$_{2}$. (b) The temperature dependent
resistance for cooling-heating sweeps of
1\emph{T}-TaS$_{1.6}$Se$_{0.4}$ crystal. The hysteresis region is
intimately correlated with the extent of the CCDW phase transition. For
the case of cooling down to 200 K followed by heating, nearly no
hysteretic feature can be identified. (c) Similar to (b), but for the
pristine 1\emph{T}-TaS$_{2}$.}
\end{figure}
\begin{figure}[b]
\centering
\begin{minipage}{0.49\linewidth}
\centering
\includegraphics[width=\textwidth]{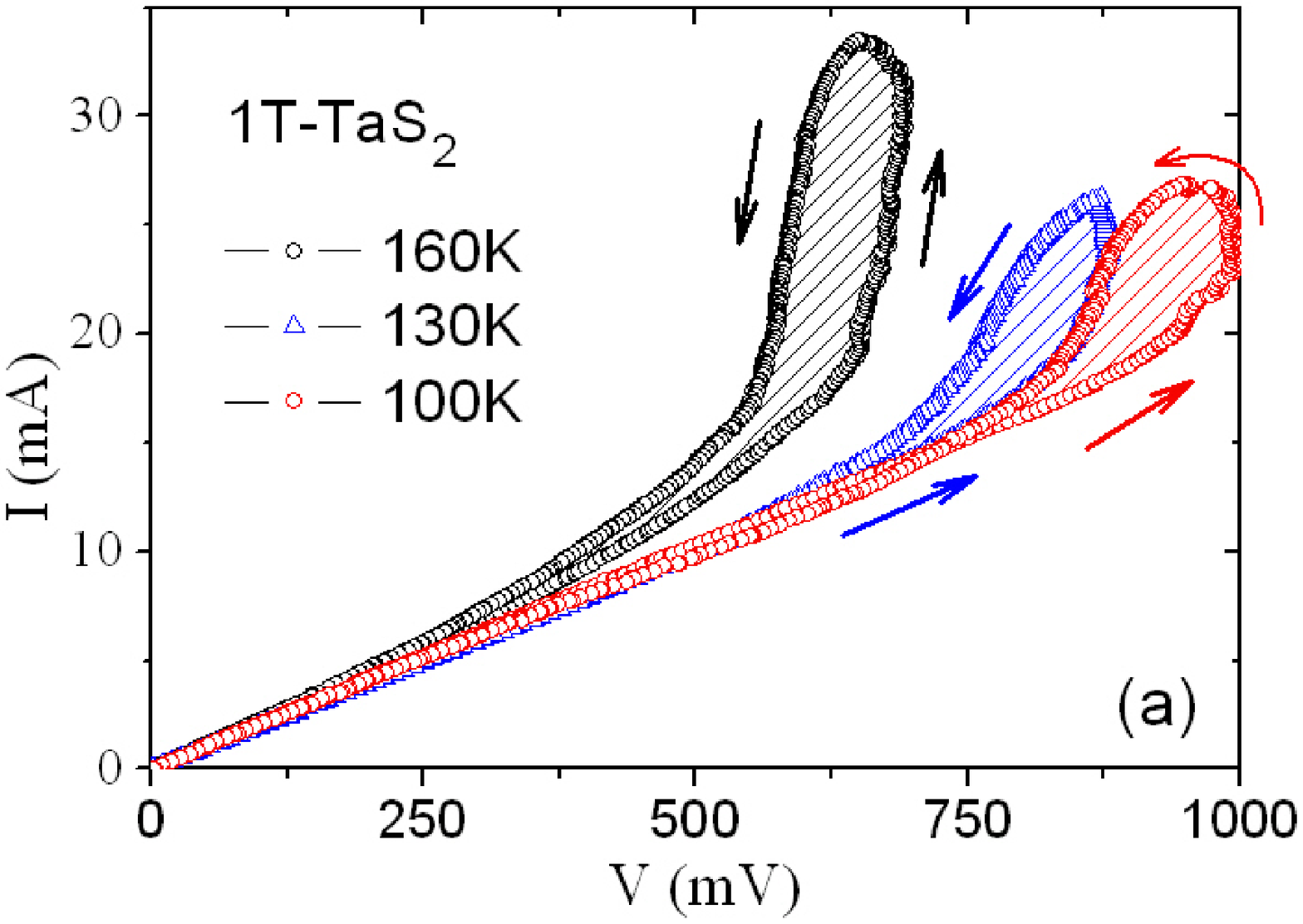}
\end{minipage}
\begin{minipage}{0.49\linewidth}
\centering
\includegraphics[width=\textwidth]{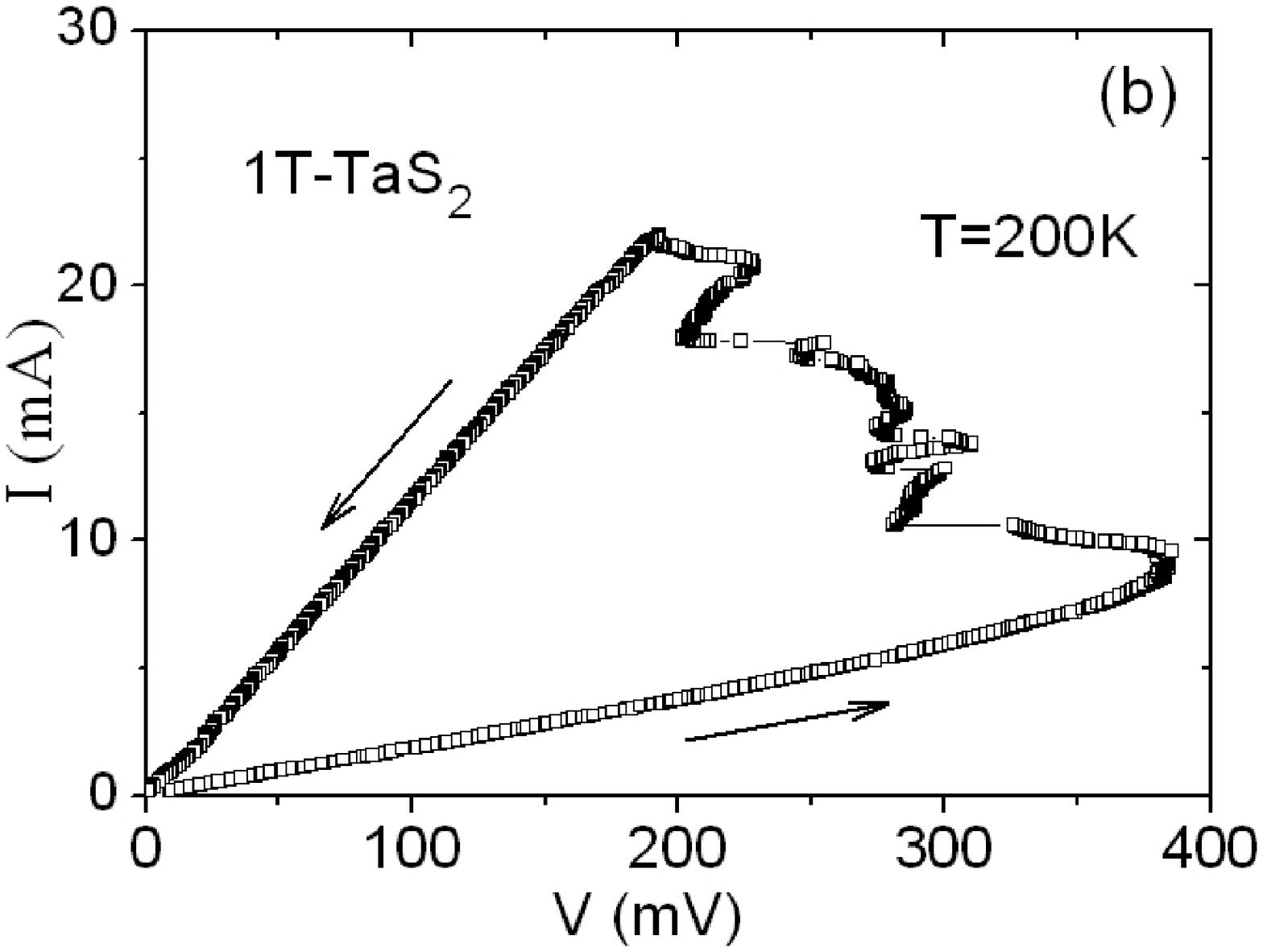}
\end{minipage}\caption{\label{Fig02-IV} (Color online)
(a) The typical current voltage characteristics of 1T-TaS$_{2}$ at 100
K, 130 K and 160 K, continuously sweeping voltage mode. The Joule
heating effect is represented by the shaded region. (b) Similar to (a),
but at \emph{T}=200 K.}
\end{figure}

The temperature dependent resistivity $\rho_{dc}$($T$), and the
hysteresis features of 1\emph{T}-TaS$_{2}$ and
1\emph{T}-TaS$_{1.6}$Se$_{0.4}$ upon warming, are consistent with
previous reports \cite{DiSalvo,Vaskivskyi,Yoshida2}, see Fig.
\ref{Fig01-IV-Vt}(a). Compared with the pristine
1\emph{T}-TaS$_{2}$, 1\emph{T}-TaS$_{1.6}$Se$_{0.4}$ exhibits a
widened hysteresis temperature range. For the hysteresis behavior
in $\rho_{dc}$($T$), it is clear that the thermal history plays
essential roles: the emergence of the high-resistance CCDW is
closely related to the hysteresis, as shown in Fig.
\ref{Fig01-IV-Vt}(b) for a 1\emph{T}-TaS$_{1.6}$Se$_{0.4}$
specimen. From the various cooling-heating sweeps of
$\rho_{dc}$($T$), the hysteresis becomes more remarkable as the
phase transition goes deeper towards the CCDW state. Specifically,
if cooling does not reach the critical temperature of the CCDW
phase transition, no identifiable hysteretic feature can be
observed. In other words, the CCDW phase is a prerequisite for the
birth of the hysteresis. In the following, the experimental data
were obtained at fixed temperatures in the hysteresis region after
warming the cryostat from a deep CCDW state at \emph{T}=30 K. Fig.
\ref{Fig02-IV}(a) clearly shows the current voltage characteristic
(CVC) of a typical 1\emph{T}-TaS$_{2}$ sample: the Joule heating
induced characteristics (by continuously voltage sweeping) appear
above 20 mA at \emph{T}=160 K, 130 K and 100 K, whereas the
primary state is recovered once the current goes towards zero.
However, at \emph{T}=200 K in the hysteresis region, a remarkable
non-volatile behavior emerges as shown in Fig. \ref{Fig02-IV}(b).

Initially, the system is in the thermodynamically stable CCDW
state with \emph{R}= 60 ohm. Upon increasing current \emph{I}, HR
state is almost independent of current until a kink appears at a
threshold $I_{th}$ = 10.0 mA. Further increase in current
decreases the voltage up to $I$ slightly exceeding 20.0 mA in the
CVC curves. On reducing current, an enhancement of the slope of
the CVC emerges and persists down to \emph{I} = 0 mA, exhibiting a
nonvolatile resistance. In contrast to the forward current scan,
the kink feature in the CVC does not appear in the backward scan,
implying the LR state survives.

The resistance near \emph{I}= 0 mA upon increasing current is
distinctly different from the case of the inverse sweeping shown
in Fig. \ref{Fig02-IV}(b), excluding the classical impurity
pinning effects. We propose that this does not originate from the
destruction of the sample, as the subsequent measurements with
several thermal and electric field sweepings reproduced the data
within experimental errors. The resistance-change effects are not
associated with dielectric breakdown into filamentary paths, as in
the memristor of a TaO$_{x}$-based asymmetric passive device
\cite{Huang,Prakash}. Zener tunnelling and avalanche breakdown can
also be eliminated, for the occurrence of metastability in our
experiments. The experimental results definitely showed that the
sample states are intimately related to the history of the
measurement, similar to the original nonvolatile switching results
reported by Yoshida \emph{et al.} \cite{Yoshida,Yoshida2} and Tsen
\emph{et al.} \cite{Tsen}. In comparison, the threshold electric
field $E_{th}$ in our experiments is remarkably smaller than the
results of Hollander \emph{et al.} \cite{Hollander}, probably due
to higher temperature of the specimen and the finite-size effects
\cite{McCarten} for the CDW systems.

\begin{figure}[b]
\centering
\begin{minipage}[t]{0.90\linewidth}
\centering
\includegraphics[width=\textwidth]{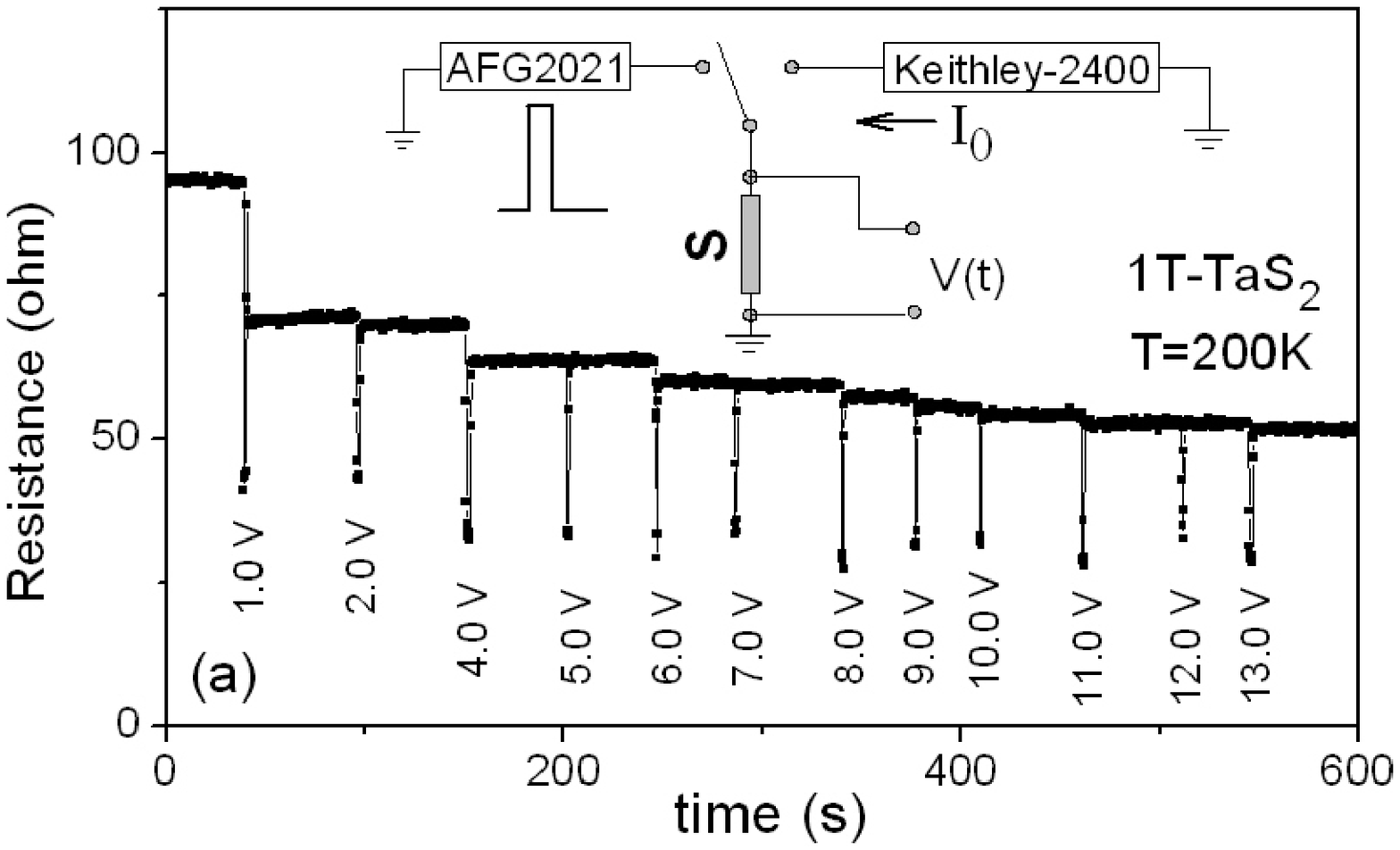}
\end{minipage}
\begin{minipage}[t]{0.48\linewidth}
\centering
\includegraphics[width=\textwidth]{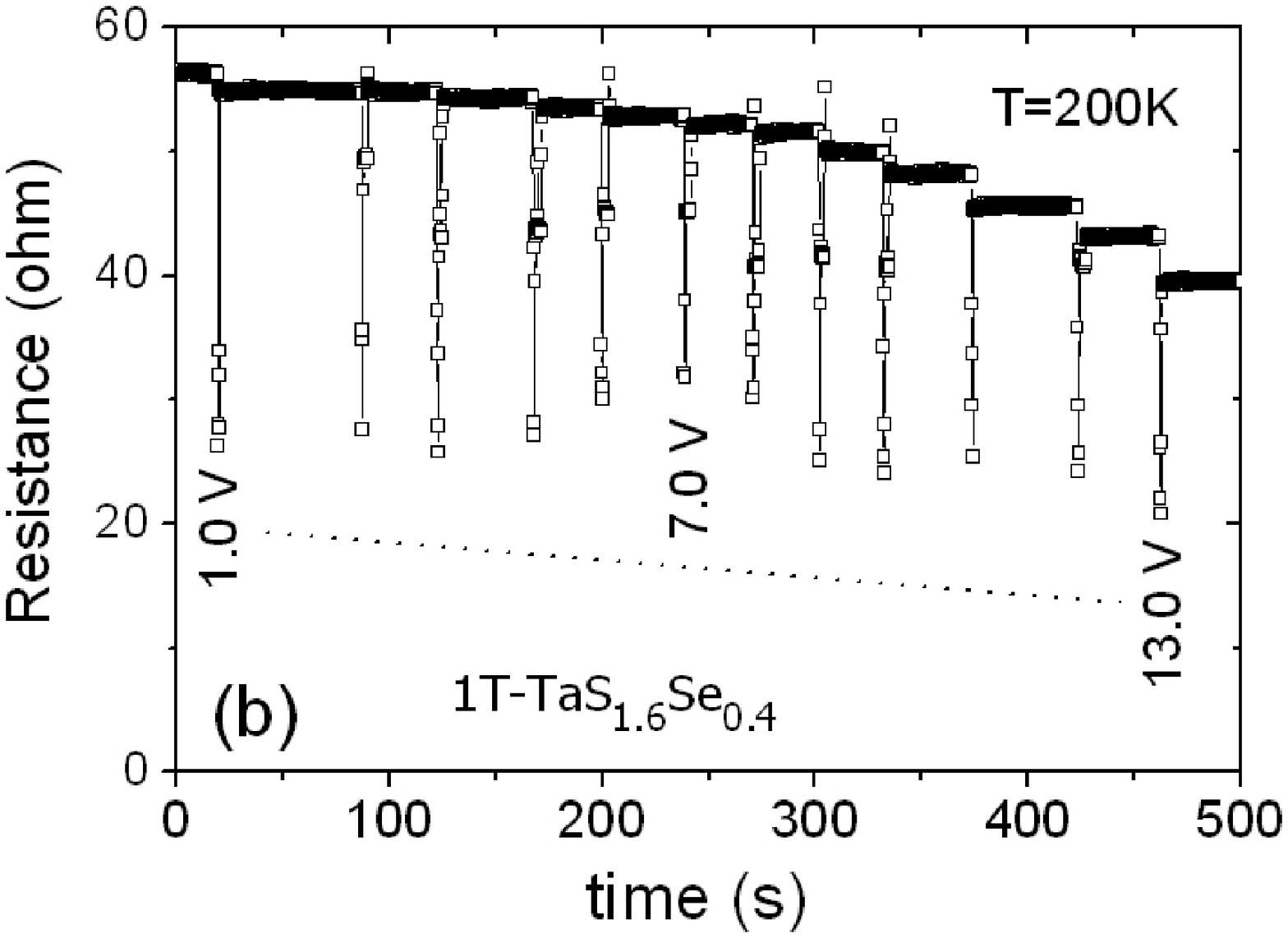}
\end{minipage}
\begin{minipage}[t]{0.49\linewidth}
\centering
\includegraphics[width=\textwidth]{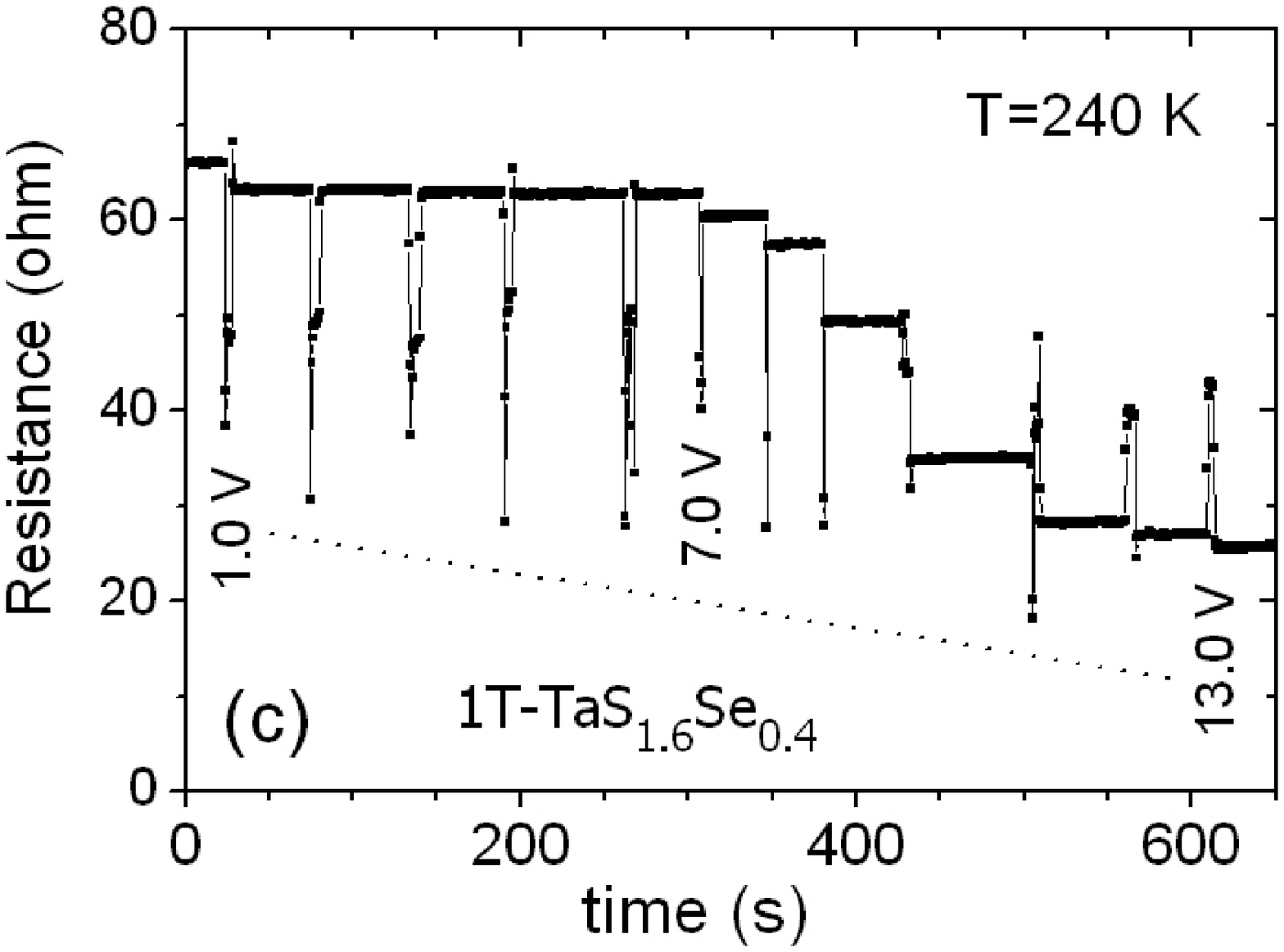}
\end{minipage}\caption{\label{Fig-Vt-200K}
(a) Time dependence of the resistance of the specimen (\textbf{S})
excited by electric pulses with duration 50-$\mu$s to avoid Joule
heating, at \emph{T}= 200 K. The inset shows the schematic circuit of
the experiments. (b) similar to (a), but for
1\emph{T}-TaS$_{1.6}$Se$_{0.4}$. (c) For
1\emph{T}-TaS$_{1.6}$Se$_{0.4}$ at \emph{T}=240 K.}
\end{figure}

The transitions from one meta-stable state to another by current
sweeping indicate that the applied electric field may have
important effects on the 1\emph{T}-TaS$_{2}$ system. To reveal the
evolution of the transport properties with increasing dc current,
we investigated the electric pulse response of 1\emph{T}-TaS$_{2}$
and 1\emph{T}-TaS$_{1.6}$Se$_{0.4}$ samples systematically. To
avoid the Joule heating, the duration of pulses should be shorter
than the relaxation time of the sample by self-heating,
$\tau_{SH}=\frac{N_{M}c}{\kappa}\cdot\frac{L}{S}$, where $N_{M}$
is the number of moles of material with specific heat \emph{c} and
thermal conductance $\kappa$ \cite{Cohen}, \emph{L} and \emph{S}
are the length and the cross-section area of the specimen,
respectively. In our experiments, it is estimated that
$\tau_{SH}\sim$ 80 $\mu$s \cite{Yongchang_2019}, so we applied
\textbf{single pulse} with duration 40 $\mu$s, the undistorted
wave form of the response indicates that the Joule-heating effects
could be neglected.

\begin{figure}[b]
\centering
\begin{minipage}{0.98\linewidth}
\includegraphics[width=\textwidth]{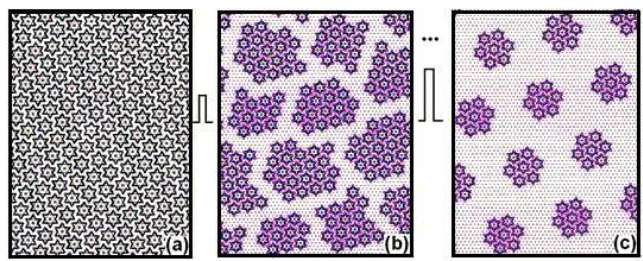}
\end{minipage}
\caption{\label{Fig04-CCDW-domain-pattern} (Color online) Real-space
CDW reordering of electronic structure. (a) Idealized diagram of CCDW
phase. (b) The schematic non-equilibrium pattern of the clusters after
an electric pulse of 1\emph{T}-TaS$_{2}$ in real space. (c) The
schematic illustration for the charge arrangements at an equilibrium
long time after an excitation of electric pulse.}
\end{figure}

At \emph{T}=200 K, a typical temperature in the hysteresis region,
we conducted the measurement using the circuit schematically shown
in the inset of Fig. \ref{Fig-Vt-200K}(a). Similar results could
also be seen in Figs. \ref{Fig-Vt-200K}(b) and
\ref{Fig-Vt-200K}(c) for Se doped sample
1\emph{T}-TaS$_{1.6}$Se$_{0.4}$. It is clear that the resistive
changes at small voltages are more significant in the pristine
crystal 1\emph{T}-TaS$_{2}$ than in
1\emph{T}-TaS$_{1.6}$Se$_{0.4}$. The associated driving force may
be from the larger $^\backprime$temperature gap$^\prime$ from the
measured temperature 200 K to the C$\rightarrow$NC CDW transition
of 1\emph{T}-TaS$_{1.6}$Se$_{0.4}$ (\emph{T}=260 K) compared with
1\emph{T}-TaS$_{2}$ (\emph{T}=220 K). The results in Figs.
\ref{Fig-Vt-200K}(b) and \ref{Fig-Vt-200K}(c) (and also our
results in Ref. \cite{Yongchang_2019}) indicate that the closer to
the C$\rightarrow$NC CDW transition the more significant of the
resistance changes. After each electric pulse, an abrupt decrease
of the resistance appears, and then maintains the state quickly
without remarkable relaxation processes in our time resolution. It
is obvious that pulses stimulate the configuration of the system
to a new meta-stable state, and the series of pulses applied
induce the multi steps of the resistance. It is believed that
further applying higher electric pulses could drive the system to
the states with larger conductivities, even the NCCDW phase. In
other words, there are multi meta-stable states between CCDW and
NCCDW phases, thus the energy differences between the meta-stable
states would be very small \cite{Tsen}. If it is the case, the
thermal energy 200 K (17 meV) would smear out the distinctions
between the multi meta-stable states, which is in conflict with
the experimental data. In Ref. \cite{Yongchang_2019}, we have
proposed a model to explain the multi meta-stable states or
namely, nonvolatile resistance behavior, induced by electric
pulses: each meta-stable state corresponds to a mixed state where
the CCDW domains are separated by metallic regions, and the
conductance of the whole system varies with the ratio of the two
components.

\begin{figure}[t]
\centering
\begin{minipage}{0.75\linewidth}
\includegraphics[width=\textwidth]{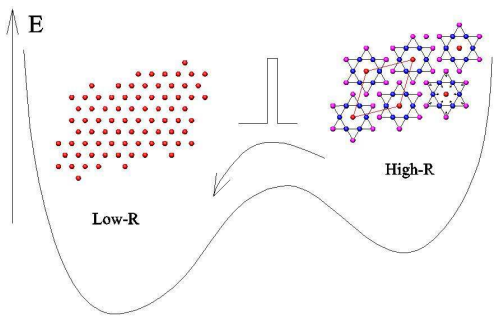}
\end{minipage}
\caption{\label{Fig05-meta-state} (Color online) The schematic
illustration we proposed for that the high conducting components are
obtained by applying electric pulses. The David-star clusters are
expected to be unbounded or liberated and thus the concentration of
free carriers proliferates, corresponding to the
$^\backprime$blank$^\prime$ regions outside of the CCDW domains in Fig.
\ref{Fig04-CCDW-domain-pattern}.}
\end{figure}

Thus we suggest the scenario for charge dynamics of
1\emph{T}-TaS$_{2}$ in the hysteresis region: in the absence of
the electric pulse, the system is in a meta-stable CCDW state [see
\ref{Fig04-CCDW-domain-pattern}(a)], consisting of the David-star
unit cells in which the electrons are bound inside, a state of HR.
Within the electric pulse duration, the CCDW phase is excited to
bear considerable strain or distortions, see Fig.
\ref{Fig04-CCDW-domain-pattern}(b). For a pulse with moderate
magnitude, the excited state is not a full NC state as it would
correspond to about 1/10 of the CCDW resistance, whereas in our
experiments the resistance jump is no more than a half as shown in
Figs. \ref{Fig-Vt-200K}(a) and \ref{Fig-Vt-200K}(b). With further
increasing the magnitudes of the electric pulses at \emph{T}=240 K
for 1\emph{T}-TaS$_{1.6}$Se$_{0.4}$ [in Fig.
\ref{Fig-Vt-200K}(c)], it appears that the resistance value comes
closer to NCCDW state as shown schematically in Fig.
\ref{Fig04-CCDW-domain-pattern}(c). Considering the system can
relax back to the equilibrium CCDW state after the activation of
electric pulses below 130 K (see Supplementary Information I),
whereas in hysteresis region the resistance value no longer
restores the initial value (the intact CCDW state), we propose
there should be some region with higher conductivities emerge in
the CCDW environment and survive. In these high-conductance
components, the David stars are expected to be unbounded/liberated
by the electric pulses and the concentration of free carriers
proliferates, forming $^\backprime$blank$^\prime$ regions outside
of CCDW domains and corresponding to the transition over the
energy barrier in Fig. \ref{Fig05-meta-state}. This carrier
proliferation process is not the same as formation of the domain
walls in the hidden state in Refs. \cite{MaLG,Karpov,Cho}, due to
the higher temperature range in our present experiments and the
relaxations after electric pulses (see Supplementary Information).

\begin{figure}[b]
\centering
\begin{minipage}{0.75\linewidth}
\includegraphics[width=\textwidth]{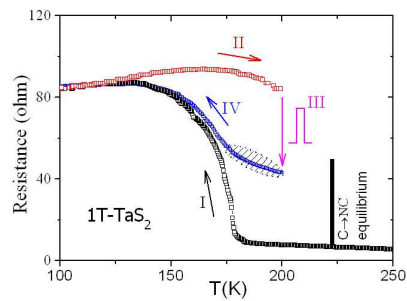}
\end{minipage}
\caption{\label{Fig06-pulse-cooling-RT} (Color online) The temperature
dependent resistance of the pristine 1\emph{T}-TaS$_{2}$ crystal. After
heating to 200 K (II), an electric pulse (10 V, 40 $\mu$s) was applied,
after which further cooling continued. The arrows I, II and IV indicate
the cooling or heating directions. Note the hatched region represented
by the ellipse, in which the curve shows an obvious enhancement of the
resistance with cooling.}
\end{figure}

The validity of our proposal is supported by the results in Fig.
\ref{Fig06-pulse-cooling-RT}. After applied electric pulses at
\emph{T}=200 K, the measured resistance shows an obvious
enhancement with cooling, and below 150 K the system reverts back
to the thermodynamically stable CCDW state, exhibiting a widened
temperature range of the transition process. In other words, the
starting point shifts towards higher temperature compared with the
thermodynamically accessible NC$\rightarrow$C phase transition.
The origin is likely the inhomogeneity of the CCDW domains induced
by pulses [see Fig. \ref{Fig04-CCDW-domain-pattern}(b)], and thus
a spread phase transition accordingly. It is expected that the
larger the domains, the higher transition temperature to the CCDW
state, as the thermal disturbance becomes insignificant. In
comparison, for the thermodynamically accessible NC$\rightarrow$C
transition, as the NC phase consists of regular CCDW domains
separated by discommensurabilities \cite{Thomson}, the phase
transition occurs homogeneously and sharply.

There are several possible explanations for the nonvolatile
electric resistance behavior. The Frenkel-Kontorova (FK) model of
the incommensurate commensurate transition for the emergence of
Devil$^\prime$s staircase \cite{Bak} is related to the numerous
minima of the free energy landscape, forming different meta-stable
states of the system. In 1D-CDW, Gruner proposed that in a
macroscopic specimen, the number of meta-stable states is about
the ratio of its scale to the coherence lengths \cite{Gruner}.
Refs. \cite{MaLG,Lee} suggest that the inter-layer stacking is a
decisive factor in determining the electronic structure of
1\emph{T}-TaS$_{2}$. In a very recent report \cite{Yoshida3},
Yoshida \emph{et al.} proposed that in the transient unsteady
state where the current is increased, the inter-layer coherence
can be lost even inside a CDW domain. As some CDW layers move an
atomically short distance ahead of the adjacent layers to settle
into a meta-stable stacking configuration, the resistance changes
are eventually established.

Though the whole mechanisms of the multi resistance states induced
by electric pulses has not been completely disclosed yet, the
potential applications are of interest in SSED. Recently, an
asymmetric passive switching device was used in Pt/W/TaO$_x$/Pt
\cite{Kim,Prakash}. Besides the function of switches by applying
appropriate voltages, the device with tunable multi-level
resistance states induced by electric pulses could be used for
synaptic devices in neuromorphic computing. We note that, in
comparison, the creation and operating for case of
1\emph{T}-TaS$_{2}$ costs much less since it is only a
two-terminal device, like a conventional resistor.

In summary, the electric pulses induced responses of
1\emph{T}-TaS$_{2}$ and 1\emph{T}-TaS$_{1.6}$Se$_{0.4}$ crystals
in the commensurate charge-density-wave phase in hysteresis
temperature range have been investigated. The Joule heating
effects were successfully avoided by the application of single
electric pulse with duration of 40 $\mu$s. Abrupt multi steps of
the resistance excited by electric pulses were observed without
remarkable relaxation processes above or at 200 K. Se doping
widens the hysteresis region of 1\emph{T}-TaS$_{1.6}$Se$_{0.4}$
crystal, and thus enhances the usage temperature of the
nonvolatile resistance properties in the layered dichalcogenides.
We propose that the response of the system may correspond to the
rearrangements of the textures of CCDW domains, The
multi-resistance states excited simply by electric pulses have
profound significance for the explorations of solid-state devices.

Data Availability Statement --- All the data that support the
findings of this study are available from the corresponding author
upon reasonable request.

Supplementary Material --- See supplementary material for
supportive data and the associated studies of the mixed state.

ACKNOWLEDGMENTS: The research work was supported by the National
Science Foundation of China (Grant No. 10704054) and Tianjin
Natural Science Foundation (No. 19JCYBJC30500). Work at Brookhaven
National Laboratory is supported by the US DOE, Contract No.
DE-SC0012704.

\end{document}